\documentclass[preprint,showpacs,amsmath,amssymb,prb]{revtex4}

\usepackage{graphicx,epsfig}
\usepackage{rotating,texdraw}

\usepackage{bm}

\begin{document}

\title{Quantum breathers in a nonlinear Klein Gordon lattice}

\author{Laurent Proville}
\email{lproville@cea.fr}
\affiliation{\small Service de Recherches
de M\'etallurgie Physique, CEA-Saclay/DEN/DMN
        91191-Gif-sur-Yvette Cedex, France}

\date{\today}

\begin{abstract}
The quantum modes of a nonlinear Klein Gordon lattice have been
computed numerically [L. Proville, Phys. Rev. B {\bf 71}, 104306
(2005)]. The on-site nonlinearity has been found to lead to a
phonon pairing and consequently some phonon bound states. In the
present paper, the time dependent Wannier transform of these
states is shown to exhibit a breather-like behavior, i.e., it is
spatially localized and time-periodic. The typical time the
lattice may sustain such breather states is studied as a function
of the trapped energy  and the intersite
lattice coupling.
\end{abstract}

\pacs{63.20.Ry, 03.65.Ge, 11.10.Lm, 63.20.Dj}

\maketitle

\section{Nonlinear lattice modes}

The discrete breather solutions are currently a matter of
intensive research (see
Refs.\cite{Siev,MA94,bishop96,Aubry,Cretegny,
mackay2000,Kopidakis,Sievers2002,
Flach2002,Fleurov2003,Eilbeck,Rosenau,Tretiak,Gomez}). The
distinctive property of those lattice modes is to gather the
spatial localization and the time periodicity so they lead to a
energy trapping and thus a delay in the
equipartition\cite{Cretegny}. As a general consequence of
anharmonicity, the emergence of breathers may be recognized as a paradigm
of physics since it occurs at different scales in various contexts, e.g., in
macroscopic networks as a chain of coupled pendulums, in
microscopic Josephson arrays\cite{Trias} as well as in
molecules\cite{benzen}, polymers\cite{Adachi} and crystals as the
PtCl ethylene diamine chlorate\cite{PtCl}.

The nonlinear excitations in materials have been studied for
several decades. In the late fifties\cite{Gush1957}, the possible
existence of a two-phonon bound state was pointed out in the
infrared (IR) spectroscopy of H$_2$ solid. The vibrational and
rotational nonlinear excitations in the H$_2$ crystal have been
thoroughly investigated both experimentally\cite{Mao} and
theoretically\cite{Kranendonk}. About the same
period\cite{Abbott}, the spectrum anomalies of the crystalline
acetanilide (ACN) was revealed and later interpreted with
different theories (see Refs.
\cite{Edler2003,Edler2004a,Edler2004b} and for a historical
survey see Ref. \cite{ScottLastChapter}). Early in the
sixties\cite{Ron}, in the HCl solid, the anharmonicity of the
first overtone of hydrogen vibration has been measured by IR
adsorption. It has been interpreted as a two phonon bound state,
namely a biphonon\cite{Gellini}, in regard of the earlier
theoretical work of V.M. Agranovich\cite{AGRA1970,AGRANO}. The
triphonon has also been identified in the spectrum of
HCl\cite{Gellini}. In seventies, similar phonon bound states have
been recognized in several molecular crystals such as CO$_2$,
N$_2$O and OCS\cite{Schettino}, as well as in water
ice\cite{RonH2O} by measuring the anharmonic IR absorptions. F.
Bogani achieved some convincing simulations of these anharmonic
spectra\cite{Bogani} by using the technics of renormalized
perturbation theory. For the last decade, the nonlinearity has
emerged in several other materials:
\begin{itemize}
\item[-] The inelastic neutron
scattering (INS) has revealed the
phonon bound
states in the metal hydrides as PdH\cite{Ross1998} or TiH and
ZrH\cite{Kolesnikov}.
\item[-] The INS has also permitted to infer proton dynamics in the molecular
crystals as polyglycine\cite{Fillaux0} and
4-methylpyridine\cite{Fillaux2}. In the latter, the bound states
of the methyl group rotational modes proved to last several days
(see Refs. \cite{Fillaux2,Fillaux1} and Ref.
\cite{ScottLastChapter} for a survey of the theory).
\item[-] The stretch overtone of carbon monoxide adsorbed on
Ru(100) has been found to exhibit a strong anharmonicity at low
surface coverage\cite{Jakob,Bonn}. Several theoretical
approaches\cite{Jakob,Bonn,Pouthier} have been attempt to analyze
the IR spectroscopy on Ru(100):CO.
\end{itemize}
The previous list is probably not complete but it is sufficient to
emphasize that the nonlinear excitations have been worked out in many
different materials, whether it is a molecular
crystal\cite{Bogani}, a hydrogen-bonded crystal\cite{Gellini} or a
metal hydride\cite{Ross1998}. Furthermore the nonlinearity may
occur in one\cite{Fillaux2}, two\cite{Jakob} and
three-dimensional\cite{Bogani} systems. In most of the above cited
examples, the phonon dispersion may be evaluated as smaller
than $10\%$ of the fundamental optical excitation and the
anharmonicity proves to reach less than $5\%$. The latter estimation
holds for the first overtone whereas for higher orders the
strength of the anharmonicity may increase as it is the case in HCl
solid\cite{Gellini} or stabilize as in PdH\cite{Ross1998}.
The
phonon bound states are the siblings of breathers as they all
stem from anharmonicity (see
Refs.\cite{ScottLastChapter,Fleurov2003} and Refs. therein).
The quantum breather may be viewed as a Wannier transform, applied
to the phonon bound states that participate to a same energy band.
Presently, the purpose of our work is to study this idea within
the nonlinear KG model.

The accurate computation of nonlinear modes, whether it is in a
classical lattice or in a quantum one requires the use of
numerics.
Recently\cite{LPXarch}, we proposed a numerical method that
permits to compute the nonlinear quantum modes in a Klein Gordon
lattice (KG) for different type of nonlinearity. In the present
paper, we use those developments to study the Wannier
transform\cite{Fleurov2003} of the lattice eigen-modes that
exhibits a quasi-particle spectrum, i.e., a narrow energy band.
The time dependant Wannier transform of these states is found to
exhibit a breather-like behavior, i.e., it is spatially localized
and time-periodic. The life time of such breather states is
studied as a function of their energy. We found that the higher
the energy spike is, the longer it remains localized. That study
has been carried out for different model parameters, including the
case where the phonon dispersion is larger than the anharmonicity.

After a brief introduction of the nonlinear KG lattice model,  our
computing method is spelled out in Sec. \ref{Sec2}. In Sec.
\ref{Sec3} and Sec. \ref{Sec33}, we present and discuss our
results on phonon bound states and breathers, respectively. Some
perspectives are given in Sec. \ref{Sec4}.

\section{Lattice model and numerical method}\label{Sec2}
\begin{figure}
\noindent
\caption{\label{fig3}  The plot of energy spectrum of a 1D chain,
composed of $N=13$ atoms for $A_4=0.2$,
versus the dimensionless coupling $C$. The eigenvalues are plotted
as empty circles excepted the phonon bound states energies, plotted
as full black circles.
The tags indicate the order of phonon bound states.}
\end{figure}

The energy of a lattice made of identical particles is expressed
as a Hamiltonian operator:
\begin{equation}
H=\sum_l [\frac{p_l^2}{2m} + V(x_l) + \sum_{j=<l>} W(x_l-x_{j})].
\label{DebyeModif}
\end{equation}
where $x_l$ and $p_l$ are displacement and momentum of the
particle at site $l$, in a one-dimensional lattice. Such a lattice
may prove relevant to model the quasi-one-dimensional networks of
quantum particles in ZrH or in PtCl. The quantum particle of mass
$m$ evolves in a on-site potential $V$, being coupled to its
nearest neighbors, $j=<l>$ by the interaction $W$. The on-site
potential $V$ is developed to the fourth order whereas $W$ is
modelled by a quadratic term:
\begin{eqnarray}
V(x_l)=a_2 x_l^2+a_3 x_l^3 +a_4 x_l^4 \nonumber\\
W(x_l-x_{j})=-c(x_l-x_{j})^2.\label{XW1}
\end{eqnarray}
Higher order terms could have been added with no difficulty for
our theory. It is possible to fixe the coefficients of $V$ within
a first principle calculation as done for PdH\cite{Elsasser} and
confirmed by the analyze of the INS spectrum\cite{Ross1998}. For
simplicity, we choose to fixe $a_3=0$. Introducing the
dimensionless operators $P_l = p_l / \sqrt{m \hbar\Omega}$,
$X_l=x_l \sqrt{m\Omega/\hbar}$ and the frequency $\Omega = \sqrt{2
(a_2-2.c)/m}$,  the Hamiltonian is rewritten as follows:
\begin{equation}
H = \hbar \Omega \sum_l \frac{P_l^2}{2} + \frac{X_l^2}{2} + A_4
X_l^4 + \frac{C}{2}  X_l  \sum_{j=<l>} X_{j} \label{Hamilton}
\end{equation}
where the dimensionless coefficients are $ A_4= a_4
\frac{\hbar}{m^2 \Omega^3}$ and  $C= \frac{4 c}{m \Omega^2}$. The
first step of our method is concerned with the exact
diagonalization of the Hamiltonian where no interaction couples
displacements. The procedure has been detailed in Ref.
\cite{LPXarch}. Arranging the on-site eigenvalues in
increasing order, the $\alpha$th eigenstate is denoted
$\phi_{\alpha,i}$ and its eigenvalue is $\gamma(\alpha)$. In case
of a negligible inter-site coupling, the $H$ eigenstates can be
written as some Bloch waves as follows:
\begin{eqnarray}
B_{ [ \Pi_i \alpha_i]} (q) = \frac{1}{\sqrt{A_{ [ \Pi_i
\alpha_i]}}} \sum_j e^{-i q.j} \Pi_i
\phi_{\alpha_i,i+j}\label{OSPBW}
\end{eqnarray}
where $A_{ [ \Pi_i \alpha_i]}$ ensures the normalization. The
label  $[\Pi_{i} \alpha_{i}]$ identifies a single on-site state
product $\Pi_i \phi_{\alpha_i,i}$ among the different products
that may be derived from the present one by translation. The set
of states $\{ B_{ [\Pi_i \alpha_i] }(q) \}_{q,N_{cut}}$, including
the uniform state $ \Pi_{i}\phi_{0,i} $ at $q=0$, form a truncated
basis where $N_{cut}$ fixes the upper boundary on the on-site
excitations: $\sum_i \alpha_i \leq N_{cut}$. In case of a non-zero
coupling, the states Eq.(\ref{OSPBW}) may be thought as some
Hartree approximation of the true eigenstates. The perturbation
theory might be applied to the intersite coupling so as to
estimate the eigenspectrum. However, we have chosen to carry out a
computation as accurate as possible. Thus the Bloch wave basis is
used to expand the Hamiltonian in. As the waves with different
$q$, are not hybridized by $H$, the Hamiltonian can be expanded
separately for each $q$. It can be achieved analytically whereas
the diagonalization of the resulting matrix has been realized
numerically with a standard method, from a numerical
library\cite{NR}. The accuracy of our calculations has been tested
both in a anharmonic\cite{LPXarch} and harmonic\cite{LPlett}
chain. For these two comparisons, a very good agreement has been
found in the two-phonon energy region and lower.
\begin{figure}
\noindent
\caption{\label{fig2mn} Energy spectrum of a 1D chain which  model
parameters are $A_4=0.2$ and $C=0.05$. The chain is
composed of $N=13$ sites. Four energy regions have been reported:
(a) phonons, (b) biphonons, (c) triphonons and (d) quadriphonons.
The eigen-energies are plotted as empty symbols and the phonon
bound states energies have been signalized by full symbols.}
\begin{picture}(300,10)(0,0)
\put(-95,330){\makebox(0,0){\Large (a)}}
\put(160,330){\makebox(0,0){\Large (b)}}
\put(-95,130){\makebox(0,0){\Large (c)}}
\put(160,130){\makebox(0,0){\Large (d)}}
\end{picture}\end{figure}

We denote by $\psi_\lambda(q)$ and $E_\lambda(q)$ the  $H$
eigenstates and the corresponding eigen-energies, respectively.
The subscript $\lambda$ fixes the correspondence between a
eigenstate and its eigen-energy. Our numerical technics allows us
to compute the scalar product $V_{\lambda,[ \Pi_i \alpha_i]}(q)$
between $\psi_\lambda(q)$ and $B_{ [ \Pi_i \alpha_i]} (q)$. Among
the Bloch waves $B_{ [ \Pi_i \alpha_i]} (q)$, we note those
bearing a single on-site excitation of order $\alpha_j>0$, all the
other lattice sites $l$, being such as $\alpha_{l}=0$. For those
states the label $[ \Pi_i \alpha_i]$ reduces to $\alpha$. In case
of $V_{\lambda,\alpha}(q)>0.5$, we choose to distinguish the
eigenstate $\psi_\lambda$ by setting $\lambda=\alpha$. It simply
means that the Bloch wave $B_\alpha (q)$ has a dominant
contribution into $\psi_\lambda (q)$. At $C=0$, one notes that
$V_{\alpha,\alpha}(q) = 1$. As it may be expected, the scalar
product $V_{\alpha,\alpha}(q)$ decreases as $C$ increases but its
variation is smooth as found in Fig.~\ref{fig3} (the results shown
in this figure are examined thoroughly in the following). The
eigenstates $\psi_\alpha (q)$ correspond to the $\alpha$ phonon
bound states. That terminology may be rightly thought  as
ambiguous since a binding energy usually refers to a groundstate
rather than to some excited states. However it is convenient as
the excitation order $\alpha$ appears in the name. This order
corresponds, indeed, to the energy level of the anharmonic on-site
potential.

\begin{figure}
\noindent
\caption{\label{fig2mnbis}
Energy spectrum of a 1D chain which  model parameters are $A_4=0.2$
and $C=0.3$. The chain is composed of either
$N=13$ sites.
Two energy regions have been reported: (a)
phonons and (b) quadriphonons.
The eigen-energies are plotted as empty symbols and
the phonon bound states energies have
been signalized by full symbols.}
\begin{picture}(300,10)(0,0)
\put(-90,125){\makebox(0,0){\Large (a)}}
\put(170,125){\makebox(0,0){\Large (b)}}
\end{picture}\end{figure}

\section{Phonon bound states}\label{Sec3}

In lattices, treated in Ref. \cite{LPXarch}, the sites
number was $N=33$ for a basis cutoff $N_{cut}=4$, which proves
sufficient for the study of the two-phonon energy region. Here, we
would like to extend our study to the case of a four phonon bound
state (quadriphonon). We thus increased $N_{cut}$ but the number
of Bloch waves, involved in our basis for $N=33$, would overload
our computer's memory, so we had to work with smaller lattices.
For $N=13$ and $N_{cut}=6$, the rank of our basis reaches $6564$
which can be managed within a reasonable time. We worked also with
a even smaller lattice, $N=7$ which allows us to increase again
$N_{cut}$ as large as $N_{cut}=9$. That case serves us as a
reference in order to test the precision of our computations on
the larger lattice.

Varying $C$ from the anti-continuous limit\cite{Aubry}, i.e.,
$C=0$  we plotted in Fig.~\ref{fig3} the eigenspectrum as a
function of $C$. Every circle symbol represents a single
eigenvalue in the half first Brillouin zone. The eigenvalues that
correspond to the eigenstates $\psi_\alpha (q)$ (described in
Sec.\ref{Sec2}) have been plotted as full circles in
Fig.~\ref{fig3}, instead of empty ones for the other eigenstates.
As far as we increased the coupling $C$ (see comment in Ref.
\cite{comment}), in Fig.\ref{fig3}, for a given order
$\alpha$ and a fixed wave vector $q$, we found a unique eigenstate
that verifies $V_{\lambda,\alpha}(q)>0.5$. This is the numerical
proof that the nonlinear excitations may be continued from $C=0$
to larger coupling. This involves that the solutions $\psi_\alpha
(q)$ conserve some features similar to the Bloch waves $B_\alpha
(q)$. Such a behavior could have been expected\cite{LPXarch} while
the energy gaps of the zero coupling spectrum remain. The
hybridization between bound and unbound phonon states is, indeed,
thought to be weak in that case. The point that was very
unexpected is that even though $C$ is large enough for gaps to
close (between the triphonon and the surrounding unbound phonon
bands, for instance) we found a dominant contribution of $B_\alpha
(q)$ into $\psi_\alpha (q)$. Moreover, for parameters in
Fig.\ref{fig3} this property does not depend on the order of the
excitation $\alpha$. It holds for phonons as for higher order
phonon bound states. What differs, however, for the latter is
their band width which increases with $C$ much smoother. In
Figs.\ref{fig2mn} (a-d), at a fixed coupling, the eigenspectrum is
plotted for different energy regions versus the wave vector. The
same symbols as in Fig.\ref{fig3} are used. We note that the
larger the energy is, the narrower the band of the phonon bound
states is. Indeed, the phonon band width is about $0.06$, whereas
for the biphonon it is less than $0.01$, for triphonon  it is
around $10^{-3}$ and quadriphonon the band width falls to
$10^{-5}$, in our energy unit. Although we approach only the very
first energy excitations, up to the fourth order, we may
reasonably extrapolate our results to higher energies. We then
expect that the band width of the phonon bound states becomes
exponentially narrower as the eigen-energy increases. In
Figs.\ref{fig2mnbis} (a-b), the coupling parameter is such as the
energy gaps close at high energy. We note that even though the
energy spectrum exhibits no gap, we find some eigenstates
$\psi_\lambda(q)$ that verify $V_{\lambda,\alpha}>0.5$ for
$\alpha=4$. In that case, the binding energy of the so called
$\alpha$ phonon bound states vanishes. However a strong component
of $B_\alpha(q)$ takes part in $\psi_\alpha(q)$. For that reason,
we propose to dub the $\psi_\alpha(q)$ eigenstates as {\it
nonlinear $\alpha$ phonons} to emphasize that these states differ
from the linear superposition of phonons, as well as to stress
their quantized feature. In Fig.\ref{fig2mnbis} (b), the band
width of the nonlinear four phonons is around $0.03$ instead of
$10^{-5}$ in Fig.\ref{fig2mn} (d). Although the width of that band
increases substantially with the coupling, it is yet one order of
magnitude below the phonon band width which is roughly $0.45$. The
exponential decrease noted at low coupling seems to be no longer
valid at larger coupling. This point deserves a thorough study
that we propose to report in a future work. In Figs.\ref{fig2mn}
(d) and \ref{fig2mnbis} (b), the band of the quadriphonon does not
exhibit the anomaly which appears when $N_{cut}$ is diminished and
that consists in a breaking of the band continuity. As far as
$C<0.3$, our numerical approach seems to be reliable to treat the
first phonon bound states. When the nonlinear parameter $A_4$ is
small, i.e., of the order of $10^{-2}$ in our
dimensionless model, the scalar product $V_{\lambda,\alpha}$ falls
below $1/2$ for a sufficiently large $C$ which depends on
$\alpha$. The larger $\alpha$ is, the larger the transition
coupling $C_\alpha$ is. Moreover the $C_\alpha$ is found to depend
on the wave vector $q$. At the edge of the lattice Brillouin zone,
$C_\alpha$ is larger than in the center. In the limit where
$A_4$ equals zero, the strictly harmonic eigenstates verify
$V_{\lambda,\alpha}=0$ as soon as $C$ is switched on. In that
particular case, $C_\alpha=0$ for all $\alpha$ but for non zero
$A_4$, the $C_\alpha$ are larger than zero, even for
$\alpha=1$ which corresponds to the single phonon.


In Ref. \cite{LPlett}, the author attempted to compare his
theoretical computations, similar to Fig.\ref{fig3}, to some
experimental measures in H$_2$\cite{Mao} solid and
Ru(100):CO\cite{Jakob}. Although such a exercise was based on
qualitative considerations, it is worthy to complete these
comparison by noting that some spectral bands are due to the
linear superposition of a biphonon and a single phonon (see
Fig.\ref{fig3} in the present paper and Fig. 3 in Ref.
\cite{LPlett}). The signature of these states has been
measured in the IR spectrum of HCl solid\cite{Gellini}. Following
a theoretical approach proposed earlier\cite{Bogani}, C. Gellini
{\it et al.} carried out the computation\cite{Gellini} of
renormalized Green functions to interpret the HCl spectrum.
Although such a theory would be inadequate for strong intersite
coupling, for molecular crystals whose the molecule's bond
anharmonicity dominates the inter-molecular coupling, as
crystalline CO$_2$ or HCl for instance, the renormalized Green
functions seems relevant to capture the main physical properties.
A convincing demonstration has been given by Bogani in Ref.
\cite{Bogani2} where a precise fit of the IR adsorption
spectrum has been achieved in several molecular crystals. The
earlier work of V.M. Agranovich introduced initially the concept
of phonon bound states and more specifically of biphonon for
interpreting some experiments where anharmonic modes had been
measured (see Ref. \cite{AGRA1970} for infrared and Ref.
\cite{AGRA1976} for neutron spectra). The lattice model of
Agranovich involves several energy terms that can be described
briefly as follows. The elementary excitation is proportional to
the on-site product of Bose-Einstein operators $a^+_{i} a_i$,
while the tunneling between neighboring sites $i$ and $j$ is
modelled by a hopping term $C a^+_{j} a_i $. The on-site Hubbard
interaction between the boson pairs simulates the lattice
anharmonicity by adding locally the energy operator $U (a^+_i)^2
a_i^2 $. Some other terms can be incorporated in the model to
modify, for instance, the biphonon tunneling\cite{AGRANO} or the
triphonon energy\cite{Kolesnikov}. These energy contributions are
parameterized by independent coefficients, e.g., $C$, $U$. For
instance, the Hubbard model for boson has been used to interpret
the INS in metal hydrides\cite{Kolesnikov}. Here the model
parameters have been adjusted to exhibit the same energy
resonances as the INS spectrum. It is possible to achieve a
similar work within the KG model, as shown in
Ref.\cite{proville2005a}. The Hubbard model for boson
involves to neglect the energy terms that do not conserve the
total boson number, despite the fact that these terms stem from
the potential energy of atoms and molecules. The consequence of
such an approximation is exemplified in computing the phonon
dispersion law. For a one-dimensional lattice, the boson Hubbard
lattice  would exhibit a phonon branch of the form $(1+C cos(q))$,
whereas the form $ \sqrt{1+2 C cos(q)}$ would be expected from the
harmonic approximation with similar parameters. The latter case
corresponds to the exact diagonalization of the linear KG
Hamiltonian which includes only the quadratic potential energy of
atoms. Since the two formula above diverge as $C$ increases, the
Hubbard model for boson proves inappropriate to treat the normal
modes at strong coupling. It might however be relevant to work out
the high order nonlinear modes as proposed in
Ref.\cite{Dubovski1994} (see also the contribution of G.P.
Tsironis in the present volume). Then the skipping of the
non-conservative boson terms might find some substantiation in the
fact that we found nonlinear narrow bands in continuous spectra as
in Fig.\ref{fig3}. A comparison between the boson Hubbard and our
KG model would be very interesting to tentatively infer the
properties of the high order phonon bound states.

\section{Quantum breathers}\label{Sec33}
\begin{figure}
\noindent
\caption{\label{fig7} Energy spectrum of a single anharmonic
oscillator versus the eigenvalue rank for different parameters:
(a) $A_4=0.2$ and (b) $A_4=0.4$. The
semi-classical calculation (empty square symbols, dashed line) is
compared to the Hamiltonian diagonalization (full circle symbols,
solid line) onto the truncated Einstein basis (see
Ref.\cite{LPXarch}) The Y axis unit is $\hbar \Omega$.}
\end{figure}
One introduces the time dependent Wannier state $W_\alpha(t,n)$,
which is constructed from a combination of the $\alpha$ phonon
bound states $\psi_\alpha (q)$. We recognize these eigenstates
among others $\psi_\lambda (q)$ by the fact that they verify
$V_{\lambda,\alpha}(q)>0.5$, for a fixed $\alpha$. This definition
permits us to build a Wannier state even though the energy
spectrum has no gap. Then the Wannier transform is written as
follows:
\begin{equation}
|W_\alpha(t,k)>= \frac{1}{\sqrt{N}}\sum_q e^{-i  (q \times k +
E_\alpha(q) \Omega t)} |\psi_\alpha (q)>. \label{WTa}
\end{equation}
The subscript $k$ indicates the lattice site where is centered the
Wannier transform. In Fig.\ref{fig3}, we found that the band of
the $\alpha$ phonon bound states contains a single state per wave
vector so that the sum over $q$ in the Wannier transform is
complete. In case of a small intersite coupling, the Bloch wave
$B_\alpha(q)$ is a good approximate of the $\alpha$ phonon bound
state with the same wave vector. This may be thought as a Hartree
approximation. To a first order in $C$, we found\cite{LPXarch}
that the $E_\alpha(q)$ dependence on $q$ is negligible provided
that $\alpha>1$ and $V$ is a single well potential. Then the
Wannier state $|W_\alpha(t,k)>$ can be rewritten as:
\begin{equation}
|W_\alpha(t,k)>=  e^{-i  ( E_\alpha \Omega t)}  \phi_{\alpha,k}
\Pi_{l\neq k} \phi_{0,i}. \label{WTa2}
\end{equation}
Such a state is localized and time periodic so it may be
considered as the quantum counterpart of the breather solutions
for the classical nonlinear discrete KG lattice. These classical
breather solutions have two important features that are first
their spatial localization and second their time periodicity with
a frequency and its overtones that are out of the linear classical
phonon branch\cite{MA94}. Our proposition could be verified by
comparing the energies of a localized time periodic Wannier state
and the semi-classical quantization of the classical breather
orbits, in same lattice. In the simple case of zero inter-site
coupling, such a comparison has been carried out in Ref.
\cite{LPXarch} and Fig.\ref{fig7}, for different on-site
potentials. The remarkable agreement allows to expect that our
proposition on breather quantum counterpart holds for larger
values of $C$. To enforce our arguments, we dwell upon
Fig.\ref{fig3}. To a fixed $\alpha>1$, the energy of the
corresponding Wannier state, given by the bracket of $H$, equals
the mean energy ${\tilde E}_\alpha$, defined as the sum of
$E_\alpha(q)/N$ over the first Brillouin zone. According to
Fig.\ref{fig3}, that mean energy does not vary much with $C$.
Indeed, provided that $\alpha>1$ and $C<0.3$ (the upper boundary
on $C$ to obtain a satisfactory precision), the bisecting line of
the phonon bound states band is roughly parallel to the X axis.
Consequently, the energy of the Wannier state of order $\alpha$ is
comparable to the same quantity computed at $C=0$. In turn, the
latter approaches very well the semi-classical quantization (see
Fig.\ref{fig7}) so the energy of the Wannier state and the one of
semi-classical breather orbits do not differ in a significant
manner provided that $C$ remains weak.  It seems reasonable in the
following to call $\alpha$ breather what is indeed the Wannier
state of order $\alpha$. It would be worth carrying out the
semi-classical quantization at non zero coupling in order to
evaluate to what extend our expectations might be confirm. We
think it should not contradict our arguments unless the classical
breather becomes unstable, i.e., its frequency or one of its
overtones fall in the spectrum of the classical normal modes.

\begin{figure}
\noindent
\caption{\label{fig4} The time evolution of kinetic energy of $17$
atoms in one-dimensional KG chain, for a Wannier state made of
phonon. The model parameters are $A_4=0.2$, $A_3=0$ and $C=0.05$.
The lattice sites are reported on the Y axis while the X axis
bears the time scale. The time unit is the inverse of $\Omega$.}
\begin{picture}(230,00)(0,0)
\put(40,150){\makebox(0,0){\Large time }}
\put(250,180){\makebox(0,0){\Large site }}
\put(-65,180){\begin{rotate}{90} {\large on-site kinetic energy}
\end{rotate}}
\end{picture}
\end{figure}

We now study the dynamics of a $\alpha$ breather as a function of
the order $\alpha$ and for a non zero coupling parameter. To that
purpose, we integrate   the time evolution of the on-site kinetic
energy $P_j^2/2$. The expectation of this operator is given by the
bracket:
\begin{eqnarray}
<W_\alpha(t,k)|\frac{P_j^2}{2}|W_\alpha(t,k)>=\frac{1}{{N}}\sum_{q,q'}
<\psi_\alpha (q')|\frac{P_j^2}{2}|\psi_\alpha (q)> \times e^{-i (
(q-q') \times k + (E_\alpha(q)-E_\alpha(q')) \Omega
t)}.\label{WTa3}
\end{eqnarray}
\begin{figure}
\noindent
\caption{\label{fig5} Profile of the 3D-plot described in
Fig.\ref{fig4} for different breathers made of: (a) phonon, (b)
biphonon, (c) triphonon and (d) quadriphonon. The parameters are
same as in Fig.\ref{fig4}. The time is reported on the X axis. The
Y axis unit is $\hbar \Omega$).}
\begin{picture}(300,10)(0,0)
\put(-85,300){\makebox(0,0){\Large (a)}} \put(160,300){\makebox(0,0){
\Large (b)}}\put(-85,100){\makebox(0,0){
\Large (c)}}\put(160,100){\makebox(0,0){
\Large (d)}}
\end{picture}\end{figure}
For $\alpha=1$, the Wannier state Eq.(\ref{WTa}) is constructed
from phonons. In the 3D plot (see Fig.\ref{fig4}) of the kinetic
energy time evolution, one notes that the energy is initially
localized and quickly spreads over the lattice. In Fig.\ref{fig5}
(a), the profile of the 3D plot shows that after $40$ time unit,
the energy is no longer localized. The same profile plot for
$\alpha=2$, in Fig.\ref{fig5} (b) shows that the quantum breather
made of biphonons may last $10$ times longer than for phonons.
With the Wannier transform of triphonons, the life time of the
localized excitation is again raised by one order of magnitude
(see Fig.\ref{fig5} (c)). The life time of the fourth quantum
breather (i.e., the Wannier transform of the quadriphonons)
overpasses the first case in Fig.\ref{fig5} (a), by three orders.
Conclusively, we found that the nonlinear KG lattice may sustain a
high energy spike for longer than $10^3$ times the typical
relaxation of low energy excitations, imposed by phonons. This
behavior is related to the dispersion of the phonon bound states
since the thinner the band is, the longer the Wannier transform
remains coherent. According to our results, the breather life time
increases exponentially with respect to $\alpha$. As noted
previously (see Sec.\ref{Sec3}), the band width decreases as the
order of phonon bound states increases, even though at high energy
the spectrum becomes continuous (see Fig.\ref{fig3}). According to
our results, this continuity does not involve a particular decay
in the breather life time.
\begin{figure}
\noindent
\caption{\label{fig6} The same as in Fig.\ref{fig5} but for
$C=0.3$ and different sizes: (a-b) $N=13$  and (c-d-e) $N=7$.  The
Wannier states are made of either (a-c) phonon, (b-d) quadriphonon
or (e) pentaphonon. The time is reported on the X axis and the Y
axis unit is $\hbar \Omega$. In the insets, some distinct time
intervals have been magnified.}
\begin{picture}(300,10)(0,0)
\put(-100,540){\makebox(0,0){ \Large (a)}}
\put(160,540){\makebox(0,0){ \Large (b)}}
\put(-100,330){\makebox(0,0){\Large (c)}}
\put(160,330){\makebox(0,0){ \Large (d)}}
\put(-20,150){\makebox(0,0){ \Large (e)}}
\end{picture}\end{figure}
It is noteworthy that the decay of a quantum breather is athermal
as it stems from the decoherence of the phonon bound states.

As $C$ increases, the $\alpha$ breather life time decreases, in
agreement to the band width enlargement shown in Fig.\ref{fig3}.
This can be worked out from the comparison of Fig. \ref{fig5} (a)
and Fig. \ref{fig5} (d) for $C=0.05$ to the left hand side insets
in Fig. \ref{fig6} (a) and Fig. \ref{fig6} (b) for $C=0.3$. Each
couple of figures concern the cases $\alpha=1$ and $\alpha=4$. For
$\alpha=1$, one sees that the life time of the localized
excitation decreases from $40$ to $4$ time units while for
$\alpha=4$, the breather life time decreases from $8\times 10^4$
to $8\times 10^1$. The drop is sharper for the higher order.
However for a fixed value of $C$, whatever this value is, the band
width of phonon bound states decreases as the energy increases so
that the life time of the corresponding breather increases too. In
Fig. \ref{fig6} (e), the life time of the fifth breather is still
two order of magnitude larger than for $\alpha=1$. We thus expect
that for a sufficiently high energy spike, the breathing mode
survives noticeably even though the intersite coupling is large.

For a fixed order $\alpha$, the breather life time does not depend
on the lattice size, as found in comparing Fig. \ref{fig6} (a) to
Fig. \ref{fig6} (c) and Fig. \ref{fig6} (b) to Fig. \ref{fig6}
(d). The Figs. \ref{fig6} (a-b) have been obtained for a $13$
sites lattice and the Figs. \ref{fig6} (c-d) for a $7$ sites
lattice. Another interesting feature revealed by these results is
the time recurrence of breather. Indeed, one notes that a certain
time after the energy spike has spread, the energy backs to its
initial trapped state, similar to the original one (see right hand
side insets in Figs. \ref{fig6} (a-e)). The breather is then bear
by few sites although it has not exactly the same amplitude as
initially. The retrapping process occurs for a time twice larger
than the breather life time because of the time inversion
symmetry. According to our computations, there is no exact
frequency for the breather recurrence as no regular behavior may
be depicted in Figs. \ref{fig6} (a-e). Though, we note that the
recurrence occurs sooner in a smaller lattice, which is
demonstrated by comparison either of Fig. \ref{fig6} (a) to (c) or
Fig. \ref{fig6} (b) to (d). In a macroscopic crystal, the
recurrence is thus expected never to take place. In contrast, the
breather recurrence might occur in a single molecule as benzen. To
that respect, the breather recurrence might be worth studying
thoroughly. Eventually, the shortest time interval upon which the
recurrence occurs after starting the breather dynamics seems to
increase with $\alpha$ as shown by comparing Fig. \ref{fig6} (c)
to Fig. \ref{fig6} (d) or Fig. \ref{fig6} (d) to Fig. \ref{fig6}
(e).

\section{Conclusion and possible developments}\label{Sec4}

As a summary, we attempted to work out
the breather modes in the
quantum KG lattice. We provided a numerical method to estimate
their life time and spatial expansion.
At the quantum scale, it proves that the breathers are closely
related to what has been called earlier, the phonon bound states
that are anharmonic eigenmodes. It is, indeed, well-known in
condensed matter physics that a narrow band excitations may be
viewed as a quasi-particle through a Wannier transform. We applied
that theory to the phonon bound states and showed that the lattice
may sustain the corresponding breather for a time which increases
as the magnitude of the energy spike. At low intersite coupling,
we found that the breather life time increases exponentially with
the trapped energy. This variation
softens at larger coupling, mainly because of the hybridization
between the phonon bound states and
the linear superpositions of lower energy modes.
For seak of simplicity, we only treated a quartic nonlinearity.
We found nonlinear excitations for all couplings we tested, i.e., up to
$C=0.3$ which corresponds to a dispersion that is larger than
the anharmonicity. In the classical counterpart of our KG
lattice, a similar result is obtained since the discrete
breathers occur at all coupling too because their frequency is higher than
the normal modes band. However,
the cubic nonlinearity is known to modify significantly this feature as
the breather frequency should be smaller than the normal modes. Consequently,
for a given breather solution, that is
for a fixed frequency, there is a coupling threshold
above which the breather is no longer stable. This transition occurs when
the breather frequency or one of its overtones fall into the classical
normal band\cite{MA94}.
A similar behavior is expected in the quantum case, which will be studied
in a future work.

The results we obtained in a KG lattice are rather encouraging for
a possible future study of the quantum acoustic lattices, as the
FPU\cite{Zab} chain.
Even though we did not address precisely that case, we found that
in energy spectra where no gap occurs, the nonlinear excitations
may yet be distinguished and still exhibit a particle-like energy
branch. Such excitations are expected to emerge in the quantum FPU
chain too, under the condition that they corresponds to a
sufficiently large energy. Our numerical theory should be
tractable on the one-dimensional FPU lattice with only few sites,
even though the rank of our basis might increase dramatically.
Then, one could yet achieve the Hamiltonian diagonalization with a
iterative procedure as the Lanczos method.

Alongside the present work, we carried out the calculation of the
dynamical structure factor of the nonlinear KG
lattice\cite{proville2005a}. A simulation of the inelastic
scattering has been achieved so as to compare our theory to
practical cases.\\




\begin{acknowledgments}
I gratefully acknowledge S. Aubry who introduced me to the theory
of breathers, at coffee breaks in Laboratoire L\'eon Brillouin
(CEA-Saclay) .
\end{acknowledgments}





\end{document}